# Nonlinear Hall effect and multichannel conduction in LaTiO$_3$/SrTiO$_3$ superlattices

J. S. Kim[1,2], S. S. A. Seo[1,3], M. F. Chisholm[3], R. K. Kremer[1], H.-U. Habermeier[1],

B. Keimer[1], and H. N. Lee[3,*]

[1] *Max-Planck-Institut für Festkörperforschung, Heisenbergstraße 1, D-70569 Stuttgart, Germany*

[2] *Department of Physics, Pohang University of Science and Technology, Pohang, Korea*

[3] *Materials Science and Technology Division, Oak Ridge National Laboratory, Oak Ridge, Tennessee 37831, USA*

We report magneto-transport properties of heterointerfaces between the Mott insulator LaTiO$_3$ and the band insulator SrTiO$_3$ in a delta-doping geometry. At low temperatures, we have found a strong nonlinearity in the magnetic field dependence of the Hall resistivity, which can be effectively controlled by varying the temperature and the electric field. We attribute this effect to multichannel conduction of interfacial charges generated by an electronic reconstruction. In particular, the formation of a highly mobile conduction channel revealed by our data is explained by the greatly increased dielectric permeability of SrTiO$_3$ at low temperatures, and its electric-field dependence reflects the spatial distribution of the quasi-two-dimensional electron gas.

Recent advances in the design of artificial transition-metal-oxide heterointerfaces [1, 2] have vastly expanded the range of materials in which electronic interface phenomena can be systematically studied and controlled. At heterointerfaces between the two band-insulators LaAlO$_3$ and SrTiO$_3$, several interesting physical phenomena have been reported, including a metal-insulator transition [3], electric-field tunable switching [3], magnetic correlations [4], and 2D superconductivity [5,6]. Since bulk transition metal oxides show a large variety of competing ground states and diverse physical properties due to strong electron correlations, incorporating such complex oxides in heterostructures provides additional opportunities for generating novel phenomena at the interface [7].

LaTiO$_3$/SrTiO$_3$ heterostructures are, in this respect, promising candidates because of several features that are distinct from those of their LaAlO$_3$/SrTiO$_3$ counterparts. The unpaired Ti $d^1$ valence electrons of LaTiO$_3$ form an antiferromagnetic Mott-insulating state due to strong electron correlations. The interfacial properties of LaTiO$_3$ are thus expected to be influenced by electronic correlations that tend to favor magnetic ground states. In fact, recent model calculations for LaTiO$_3$/SrTiO$_3$ interfaces proposed a magnetic and orbital order distinct from that of bulk LaTiO$_3$ in a wide range of parameter space [7,8], which still requires experimental verification. Moreover, both SrTiO$_3$ and LaTiO$_3$ share a common constituent, the TiO$_2$ layer. When these two materials are stacked together at the atomic scale, this produces an electron-type (TiO$_2$)$^0$/ (LaO)$^+$ interface. Importantly, Ti $d^1$ states in LaTiO$_3$ can readily accommodate additional holes, so that direct charge transfer from this state to the SrTiO$_3$ $d^0$ state can occur at the interface [1]. Even a single LaTiO$_3$/SrTiO$_3$ interface can therefore be conducting.

In order to realize electronically-reconstructed LaTiO$_3$/SrTiO$_3$ heterostructures, we fabricated various superlattices composed of unit-cell-thin LaTiO$_3$ layers epitaxially embedded in a SrTiO$_3$ matrix(Fig. 1(a)), analogous to the delta-doping geometry in conventional semiconductor heterostructures [9]. Well-defined oscillations of the reflection high-energy electron diffraction (RHEED) pattern (Fig. 1 (b)) and pronounced superlattice peaks from x-ray diffraction scans [9] as well as the sharp interfaces in Z-contrast scanning transmission electron microscopy images (Figs. 1(c), (d)) clearly confirm that the unit-cell-thin $LaTiO_3$ layers are sharply confined within the $SrTiO_3$ matrix.

The temperature dependence of the two-dimensional resistance ($R_{2D}$) for various superlattices demonstrates the presence of conducting interfaces (Fig. 2(a)), consistent with recent results from photoemission [10] and optical spectroscopy [11]. As to the origin of such metallic interfaces, one can consider two possible scenarios; (1) interfacial charge carrier doping driven by an electronic reconstruction, and (2) growth-induced extrinsic chemical doping, *e.g.* off-stoichiometry of oxygen. While still controversial, the growing consensus is that extrinsic doping by oxygen vacancies can be avoided by fabricating superlattices in an oxygen-rich environment and/or post-annealing at a high oxygen pressure [12-17]. Such systems have revealed a 2D carrier density ($n_{2D}$) of about $10^{14}$ cm$^{-2}$, and weak temperature dependence of $R_{2D}(T)$ [4,12,16]. As one can see in Figs. 2(a) and (b), our $LaTiO_3/SrTiO_3$ superlattices indeed show these features. Furthermore, neither $R_{2D}$ nor $n_{2D}$ for our superlattices exhibit any dependence on the global cation stoichiometry, *i.e.* the La fraction, $x$ = La/(La+Sr). In particular, $n_{2D}$ exhibits only a narrow range of values, unlike $(La,Sr)TiO_3$ alloys where it depends strongly on $x$ [18,19]. Based on this evidence, we therefore rule out a major influence of oxygen vacancies and/or chemical mixing on the electronic properties of our superlattices.

To reveal the nature of metallic conduction of the interfacial charge carriers, we have investigated the low-temperature transport properties, in particular, the magnetic field ($H$) dependence of the Hall resistance $R_{xy}(H)$. Figures 2(c) and (d) show $R_{xy}(H)$ at different temperatures for [L1/S10] and [L1/S6] superlattices. $R_{xy}(H)$ is negative and depends linearly on $H$ above 50 K, as found in conventional metals. At lower temperatures, however, $R_{xy}(H)$ exhibits a pronounced nonlinearity: -$R_{xy}(H)$ rises rapidly with increasing $H$, shows a broad hump in some cases, and finally crosses over to linear $H$-dependence at high magnetic fields. A similar behavior has also been observed in other superlattices [9]. Since such a strong nonlinearity in $R_{xy}(H)$ has not been observed in bulk $LaTiO_3$ nor in $SrTiO_3$, these observations suggest a novel interfacial state in $LaTiO_3/SrTiO_3$ heterostructures.

At first glance, it is tempting to consider the anomalous Hall effect (AHE), often found in ferromagnetic metals [20], as the origin of this behavior. In fact, the AHE has recently been invoked to explain transport anomalies in related systems such as Cr-doped bulk (La,Sr)TiO$_3$ [21] and LaVO$_3$/SrTiO$_3$ heterostructures [22]. Several of our observations are, however, inconsistent with this interpretation. First, we have not observed any magnetic hysteresis, which would be expected if ferromagnetism were present [9]. Second, if the effects were due to spin textures at the interface of LaTiO$_3$, as suggested for LaVO$_3$/SrTiO$_3$ superlattices [22], it would be natural to expect that the characteristic temperature $T_0$, where the nonlinearity of $R_{xy}(H)$ becomes significant, is strongly dependent on the carrier density. In bulk LaTiO$_3$, for instance, the Néel temperature $T_N$ ~140 K is dramatically reduced to zero at a doping level of a few % [18,19]. However, $T_0$ of our superlattices is almost independent on $n_{2D}$ over a rather wide range [9]. These findings have prompted us to consider an alternative origin of the non-linear $R_{xy}(H)$, even though we cannot completely rule out the possibility of interfacial magnetism due to orbital and/or charge ordering [7,8].

A nonlinear Hall effect can also arise from a multichannel conduction mechanism involving different electronic bands and/or spatially separated parallel conducting channels [20]. Although it is usually much weaker than what we observed in our LaTiO$_3$/SrTiO$_3$ heterostructures, a nonlinear Hall effect has been found in delta-doped semiconductors [23]. For high doping concentration, multiple subbands inside the potential well due to the delta doping are occupied and are involved in the charge conduction. The mobility for each subband is quite sensitive to the spatial distribution of carriers; the wave function of the lower (higher) sub-bands experiences higher (lower) scattering because of larger (smaller) overlap with the dopant layer. When only two contributions to conduction are taken into account, the $R_{xy}(H)$ can be written as $R_{xy}(H) = [(\mu_1^2 n_1 + \mu_2^2 n_2)+(\mu_1\mu_2 B)^2(n_1+n_2)]/e[(\mu_1|n_1|+\mu_2 |n_2|)^2 + (\mu_1\mu_2 B)^2(n_1+n_2)^2]$. Based on this equation, we have fitted the $R_{xy}(H)$ data with the constraint of $R_{xx}(0)) = 1/e(n_1\mu_1+n_2\mu_2)$. The strong nonlinearity of $R_{xy}(H)$, including in particular the characteristic features at low magnetic fields, are well captured by the fit of $R_{xy}(H)$ (see Fig. 2), suggesting that a similar mechanism is at work in oxide heterostructures [24].

What is unique in the LaTiO$_3$/SrTiO$_3$ system, compared to conventional semiconductor heterostructures, is the strong temperature dependence. Figure 3 shows the results of the temperature dependent fitting parameters. The fit implies a large difference in both the density and the mobility of majority ($n_1$, $\mu_1$) and minority ($n_2$, $\mu_2$) carriers. For all samples, the density $n_1$ of the majority carriers is almost independent on temperature, while their mobility $\mu_1$ increases upon lowering the temperature. On the other hand, the minority carrier density $n_2$ exhibits strong temperature dependence, and it appears too small (< $10^{10}$ cm$^{-2}$) to be detected by our Hall measurements at high temperatures. Below 50 K, however, it starts to grow rapidly by almost an order of magnitude before saturating at low temperatures. The temperature dependence of $\mu_2$ is similar to the one of $\mu_1$, but its magnitude is $10^2$ times greater, reaching 1000-5000 cm$^2$V$^{-1}$s$^{-1}$ at low temperatures.

Moreover, such a strong enhancement of the mobility of the minority carriers cannot be simply explained by usual thermal effects, *e.g.*, reduced electron-phonon scattering at low temperatures. The increasing minority carrier density, $n_2$, at low temperatures is also quite unusual. In the aforementioned model, the minority carrier density depends on how far the wave function is extended on either side of the potential well. A rough estimate [23] of the spatial extent, $z$, for the ground state wave-function in a V-shape quantum well is $z \sim (\hbar^2\varepsilon/e^2 N_D^{2D} m^*)^{1/3}$. Here, $N_D^{2D}$ is the density of dopants and $m^*$ is the effective mass of the carriers, which are usually temperature-independent. On the other hand, the characteristic width of the carrier distribution also depends on the strength of potential screening, i.e. the dielectric permittivity $\varepsilon$. In contrast to conventional semiconductors, $\varepsilon$ of SrTiO$_3$ increases by two orders of magnitude at low temperatures due to incipient ferroelectricity [25]. With the enhanced $\varepsilon$ at low temperatures, the screening of the induced electric field introduced by delta-doping becomes more effective, and the wedge-shaped potential becomes shallower, as illustrated in the inset of Fig. 3. Accordingly, the weight of charge distribution away from the delta-doping layer increases, effectively leading to an increase in the minority carrier density. The increase in the minority carrier density of our superlattice is indeed quite similar to the temperature dependence in $\varepsilon$ for bulk SrTiO$_3$ [25].

Further evidence for the close relationship between the carrier mobility and its spatial distribution is confirmed by investigating the effect of a gate voltage ($V_g$). Figures 4(a) and 4(b) show that the variation of $V_g$ induces a large modulation of both $R_{xy}(H)$ and $R_{2D}$ measured from a [L2/S6] superlattice. For a large negative $V_g$, corresponding to the more resistive state, $R_{xy}(H)$ shows approximately linear dependence on the magnetic field. As the gate voltage increases towards positive biasing, $R_{2D}$ is gradually reduced (Fig. 4(b)), and the nonlinearity of $R_{xy}(H)$ becomes remarkably stronger. The crossover from linear to nonlinear behavior of $R_{xy}(H)$ with increasing $V_g$ resembles the temperature dependence of $R_{xy}(H)$ (Fig. 2), suggesting that the two-channel model can be applied here as well. Indeed, fits of Eq. (1) to the data at different $V_g$ again yield good agreement [Fig. 4(a)]. The gate voltage dependence of the transport parameters extracted from these fits is presented in Figs. 4 (d) and (e). While the density $n_1$ and mobility $\mu_1$ of the majority carriers do not depend very much on $V_g$, the corresponding quantities of the minority carriers exhibit much stronger dependence on $V_g$. In particular, $n_2$ increases by almost an order of magnitude at high positive $V_g$, and $\mu_2$ is significantly enhanced up to 5000 cm$^2$V$^{-1}$s$^{-1}$. For negative $V_g$, on the other hand, $n_2$ quickly becomes negligible.

In the framework of the model discussed above, the origin of this behavior is readily explained as the confluence of two factors: First, the dielectric constant of SrTiO$_3$ is known to decrease in an external electric field [25]. The applied electric field reverses the increase of $\varepsilon$ at low temperatures noted above, thus reducing $n_2$ for negative $V_g$. For positive gate voltage, this effect is offset by a second factor, namely the flattening of the potential profile towards the gate electrode, as illustrated in Fig. 4(c). The shallower potential profile shifts the electron wave-function away from the scattering centers in the delta-doping layer, thereby enhancing the density and mobility of the minority carriers Moreover, considering our superlattice geometry with a maximum 2-3 nm (6 unit cells) separation between two adjacent LaTiO$_3$ layers, we expect that most of the minority carriers with a high mobility reside near the first interface from the SrTiO$_3$ substrate, because the electric field applied through the back gate affects dominantly the first interface with the substrate (see Ref. 9 for more details). These results clearly demonstrate that the observed two-channel transport in our LaTiO$_3$/SrTiO$_3$ heterostructures reflects the spatial extent of the quasi-2D electron gas, which can be modulated by the external electric field.

Previously, the spread of charge carriers at room temperature has been reported to be approximately three unit cells (~ 1 nm) [1]. Based on our results, it is expected to be wider at low temperatures due to the drastic increase of $\varepsilon$ for $SrTiO_3$. For $LaAlO_3/SrTiO_3$ heterostructures, a penetration depth of carriers, *i.e.* 10 nm at 10 K, has been estimated [26]. However, we note that unlike $LaAlO_3$, the $LaTiO_3$ layer is a $d^1$ charge reservoir and can accommodate extra charges. The exact spatial distribution of the carriers and their redistribution under electric fields need to be further investigated, especially for the highly mobile conduction channel.

In summary, we have shown how the transport properties of the 2D electron gas at an interface of $LaTiO_3/SrTiO_3$ can be modulated by delta-doping and an external electric field. Although the effect of electronic correlations in our superlattices is suppressed by a leakage of the Ti $d^1$ electrons of $LaTiO_3$ into $SrTiO_3$, diminishing the tendency towards magnetism, our finding provides new perspectives for wave-function engineering in oxide heterostructures. We also envision that bulk-like antiferromagnetism can be recovered by increasing the thickness of the $LaTiO_3$ layers, which may add another degree of freedom to manipulate the intriguing electronic properties of the electronically reconstructed interface.

We thank K. B. Lee and S. Okamoto for useful discussions and comments. The work at ORNL was supported by the Division of Materials Sciences and Engineering, U.S. Department of Energy. We also acknowledge support by the DFG under grant SFB/TRR 80.

[*]E-mail: hnlee@ornl.gov

Figure Captions

**Fig. 1** (Color online) (a) Schematic representation of a delta-doped $LaTiO_3/SrTiO_3$ heterostructure and its energy band diagram. The superlattices are denoted as [L$m$/S$n$], where L(S) refers to $LaTiO_3$ ($SrTiO_3$), and $m(n)$ indicates its thickness in unit cells. (b) RHEED oscillations during the growth of a [L1/S6] superlattice. (c) Cross-sectional Z-contrast image of a test sample with various [L$m$/S$n$] superlattices (d) High-resolution Z-contrast image of a [L2/S6] superlattice (left) and its colorized version (right).

**Fig. 2** (Color online) (a) Temperature dependence of 2D resistance per interface $R_{2D} = R_{sheet}N_{IF}$ for various configurations, where $R_{sheet}$ and $N_{IF}$ are the sheet resistance and the number of interfaces, respectively. (b) The carrier density per interface, $n_{2D} = n_{sheet}/N_{IF}$, at $T = 100$ and 280 K as a function of La fraction where the sheet carrier density $n_{sheet} = -1/R_H e$. For comparison, we also plotted the carrier density obtained from optical spectroscopy (Ref. [11]), the ideal value of the induced charge, $0.5e$ per unit cell (solid line), which is nominally expected to be transferred across the interfaces, and the charge density (dashed line) for $(La,Sr)TiO_3$ solid solutions (Ref. [18,19]). (c),(d), Magnetic field dependent Hall resistance, $R_{xy}(H)$ at various temperatures for [L1/S10] and [L1/S6] superlattices. The solid lines are fitted curves using the two-carrier model. The dashed lines indicate the result of a linear fit of the high field $R_{xy}(H)$ data.

**Fig. 3** (Color online) The temperature dependence of sheet carrier densities $n_i$ and mobilities $\mu_i$ ($i =1,2$) assuming a two-channel conduction of charge carriers in various $LaTiO_3/SrTiO_3$ superlattices. The majority charge carriers (open circles) have lower mobility, while the minority charge carriers (solid circles) have much higher mobility. At high temperatures, where $R_{xy}(H)$ shows almost linear field dependence (Fig. 2), we estimate the density and the mobility (open squares) for one type of charge carriers, which presumably corresponds to the majority charge carriers. Schematics of the potential profile and wave function of charge carriers in a superlattice are shown in the inset. The LDHM (the area in yellow) and HDLM carriers (in red) stand for the low-density-high-mobility and high-density-low-mobility carriers, respectively. The higher dielectric constant $\varepsilon$ of $SrTiO_3$ at low temperatures (left) makes the potential profile much shallower than that at high temperatures (right).

**Fig. 4** (Color online) (a) Hall resistance of a [L2/S6] superlattice as a function of magnetic field at $T$ = 3 K for various gate voltages increasing with a 20 V step between -200 and 200 V. The gate electric field was applied across the 0.5 mm-thick $SrTiO_3$ substrate through backside contacts. The solid lines are fitted curves using the two-carrier model. (b) Gate voltage dependence of the 2D resistance. (c) Schematics of the effects of external electric field on the potential profile and wave function of charge carriers in a delta-doped superlattice. Here the gate electrode is assumed to be placed on the left side of the delta-doped layer. (d),(e), The gate voltage dependence of carrier densities and mobilities for the two-channel conduction.

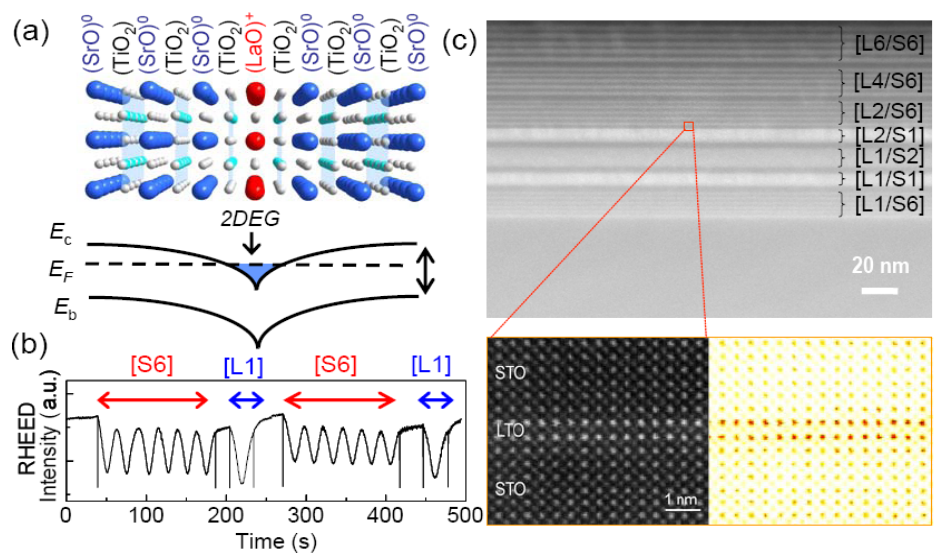

Fig. 1. Kim et al.

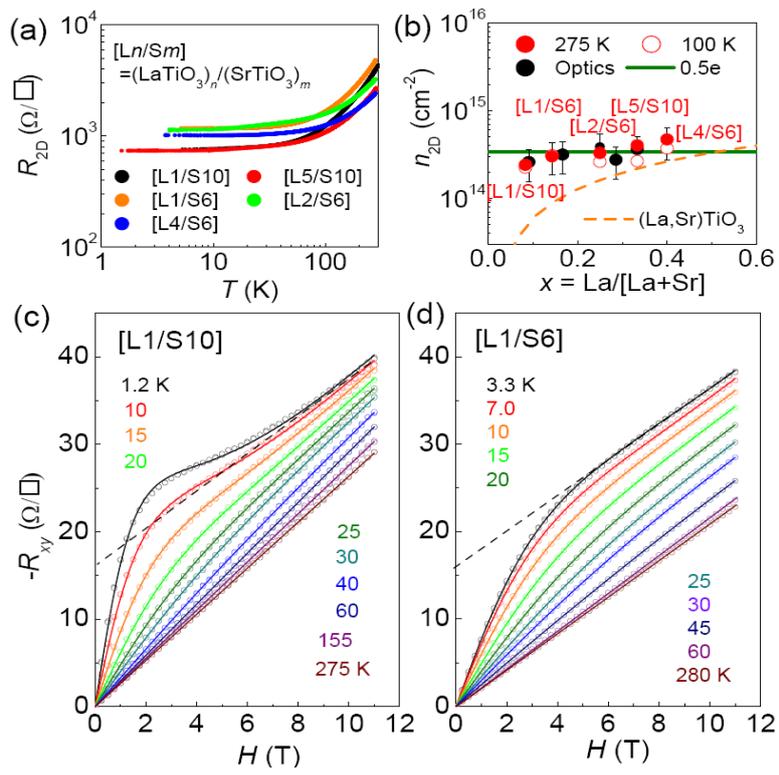

Fig. 2. Kim et al.

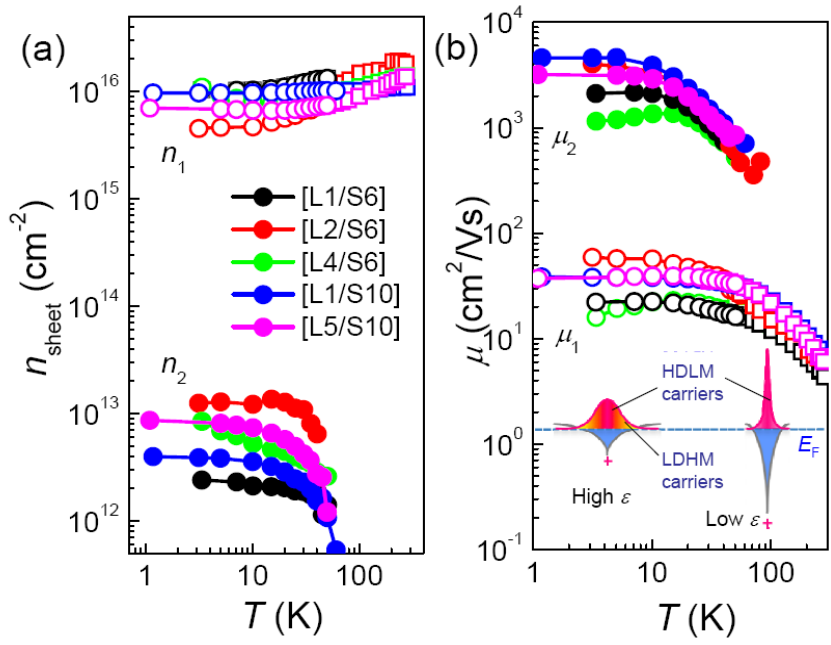

Fig. 3. Kim et al.

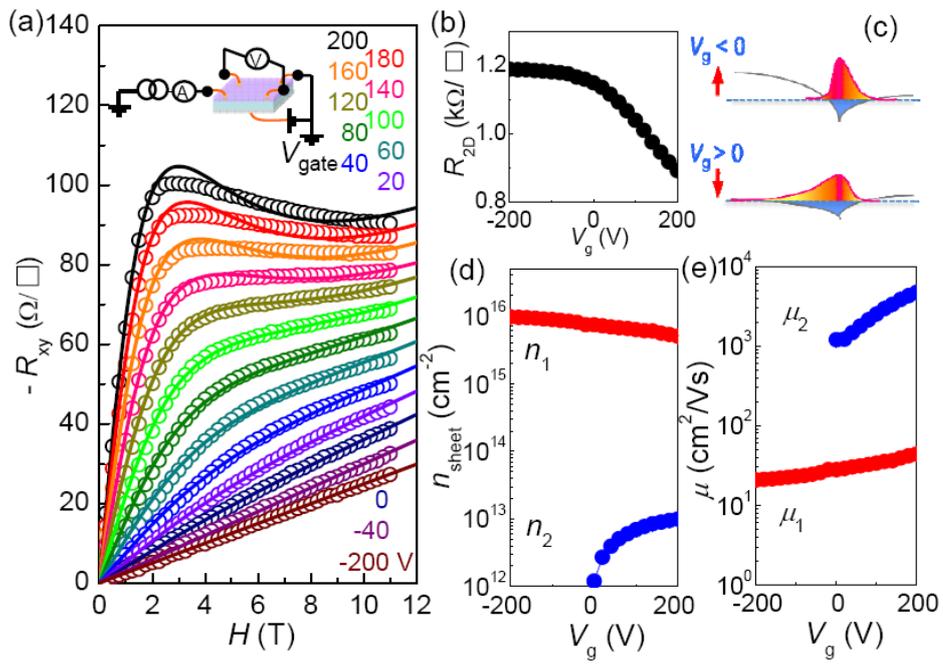

Fig. 4. Kim et al.

Supplementary Information

# Nonlinear Hall effect and multichannel conduction in LaTiO$_3$/SrTiO$_3$ superlattices

J. S. Kim[1,2], S. S. A. Seo[1,3], M. F. Chisholm[3], R. K. Kremer[1], H.-U. Habermeier[1], B. Keimer[1], and H. N. Lee[3]

[1] *Max-Planck-Institut für Festkörperforschung, Heisenbergstraße 1, D-70569 Stuttgart, Germany*

[2] *Department of Physics, Pohang University of Science and Technology, Pohang, Republic of Korea*

[3] *Materials Science and Technology Division, Oak Ridge National Laboratory, Oak Ridge, Tennessee 37831, USA*

## 1. Growth of LaTiO$_3$/ SrTiO$_3$ superlattices

Atomic-scale layering of LaTiO$_3$ and SrTiO$_3$ was done at 720 °C in 10$^{-5}$ Torr of oxygen by pulsed laser deposition using single crystal (SrTiO$_3$) and sintered (La$_2$Ti$_2$O$_7$) targets. A KrF excimer laser ($\lambda$ = 248 nm) was used for ablation of target materials at a laser fluence of ~1 J/cm$^2$. Very pronounced oscillations of reflective high-energy electron diffraction (RHEED) specular spot intensity confirmed the layer-by-layer growth with atomistic precision. After growth, all samples were exposed immediately to a higher oxygen pressure ($P_{O2}$ ~ 10$^{-2}$ Torr) for *in situ* post-annealing at the growth temperature for 5 - 30 min in order to ensure full oxidation of superlattices, followed by cooling in the same pressure. As shown in Fig. S1, high crystalline films are fully strained on SrTiO$_3$ substrates with designed in periodicities. The high quality of superlattices are also confirmed by imaging atomically flat surfaces with single terrace steps (0.4 nm in height) by atomic force microscopy (AFM). Note that the AFM image (image size: 5×5 μm$^2$) shown in Fig. S1(c) is taken from a 100-nm-thick superlattice [L1/S10].

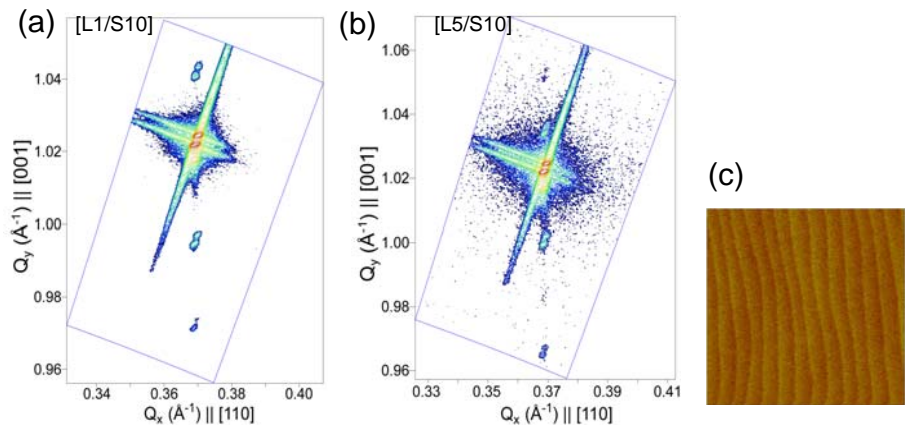

Figure S1. Reciprocal space maps of x-ray diffraction for (a) [L1/S10] and (b) [L5/S10], respectively. Note that the in-plane lattice constants of the superlattices are fully strained to those of the SrTiO$_3$ substrates. (c), Atomic force microscopy image (5×5 μm$^2$) for the surface of a [L5/S10] superlattice showing clear terrace steps of one unit-cell (~0.4 nm) height.

## 2. Magnetotransport of other LaTiO$_3$/ SrTiO$_3$ superlattices

Electric contacts were made by Ar-ion etching the lithographically defined holes of ~ 350 μm in diameter and then filling these holes by *ex-situ* Au deposition at the corners of the samples in a Van-der-Pauw geometry. This allows us to access the buried interface and measure its transport properties. No further patterning of the sample was done to avoid possible damage at the interface. The gate electric field was applied across the 0.5 mm-thick SrTiO$_3$ substrate through backside contacts. Gate leakage currents are less than 1nA for all the gate voltages between -200 V to 200 V.

The magnetic field dependent Hall resistivity of LaTiO$_3$/SrTiO$_3$ superlattices with other configurations, [L5/S10], [L2/S6], and [L4/S6] is shown in Fig. S2. Similar to the samples, [L1/S10] and [L1/S6] in Fig. 2, the strong non-linear dependence of the Hall resistivity on the magnetic field is obtained at low temperatures. Such a nonlinear field dependent Hall resistivity looks quite similar to anomalous Hall effect (AHE) that is often found in ferromagnetic metals [1]. The phenomenological expression describing the AHE consists of two terms, $R_{xy} = R_s B + R'_{xy}$, where $R_s$ is the ordinary Hall coefficient related to the carrier density and $R'_{xy}$ is the anomalous Hall resistance due to magnetic effects. Usually, $R'_{xy}$ is proportional to the magnetization $M$, and can thus serve as an alternative means to probe the field-dependent magnetization $M(H)$ when direct measurements are not feasible.[2] As shown in Figs. 2 and Fig. S2, we have estimated $R'_{xy}$ by a linear fit of the high field $R_{xy}(H)$ data. For all samples, $R'_{xy}(T)$ rises significantly as the temperature decreases [Fig. S3(a)]. In this picture, the characteristic temperatures where $R'_{xy}(T)$ begins to increase ($T_{onset}$) or show the maximum derivative ($T_{kink}$) can be viewed as manifestation of an underlying magnetic phase transition.[Fig. S3(b)]

As discussed in the text, if the effect were due to a complex interface magnetic state of LaTiO$_3$, as suggested for LaVO$_3$/SrTiO$_3$ superlattices,[3] the characteristic temperatures $T_{onset}$ or $T_{kink}$ would be quite sensitive to the carrier density. In bulk LaTiO$_3$, for instance, the antiferromagnetic state is dramatically suppressed by a few percent of doping as shown in Fig. S3(b). In contrast, $T_{onset}$ and $T_{kink}$ of our LaTiO$_3$/SrTiO$_3$ superlattices are almost independent on the carrier density over a relatively wide range of carrier concentrations. In addition, we have not found any evidence of magnetic hysteresis that is a common feature of the ferromagnetic state. Figure S4 shows the Hall resistivity and the magnetoresistivity of (LaTiO$_3$)$_1$/(SrTiO$_3$)$_{10}$ at $T = 1.1$ K with a fine step of the applied magnetic field, 300. If the strong nonlinear behaviour of the Hall resistivity as a function of the magnetic field results from a ferromagnetic origin, the anomalous Hall effect as well as the magnetoresistivity are expected to show the magnetic hysteresis as the magnetization does in ferromagnets. Note that such a magnetic hysteresis has been taken as experimental evidence for the magnetic ground state of the LaAlO$_3$/SrTiO$_3$ interface grown under high oxygen pressure conditions.[4] In our LaTiO$_3$/SrTiO$_3$ superlattice, however, we find no evidence of the magnetic hysteresis in the Hall effect nor in the magnetoresistivity down to 1.1 K.

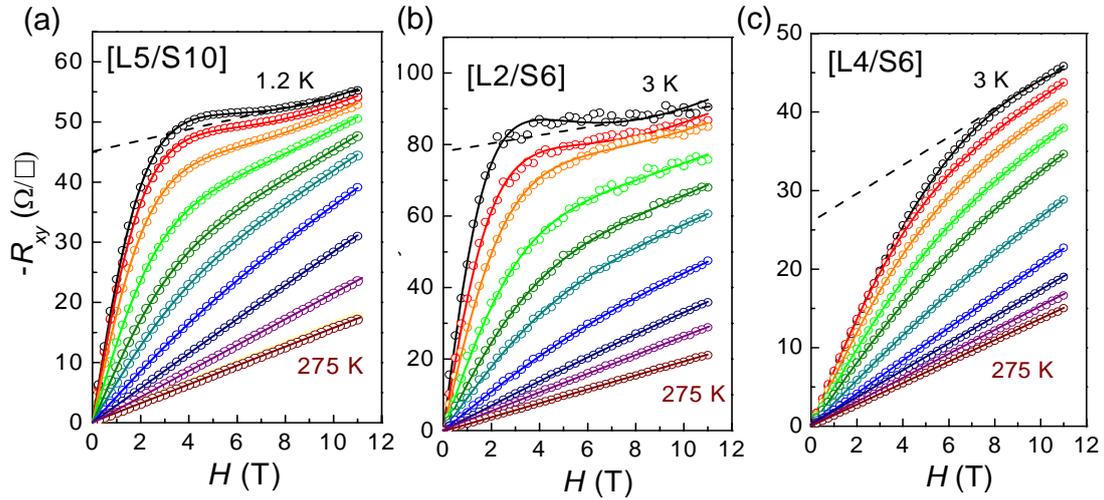

Figure S2. Magnetic field dependent Hall resistance, $R_{xy}(H)$ at various temperatures from 275 to 1.2 K for LaTiO$_3$/SrTiO$_3$ superlattices with the [L5/S10], [L2/S6] and [L4/S6] configurations. The dashed lines are the linear fit of the high field $R_{xy}(H)$ data, and the solid lines are the fitted curves using the two-carrier models described in the text.

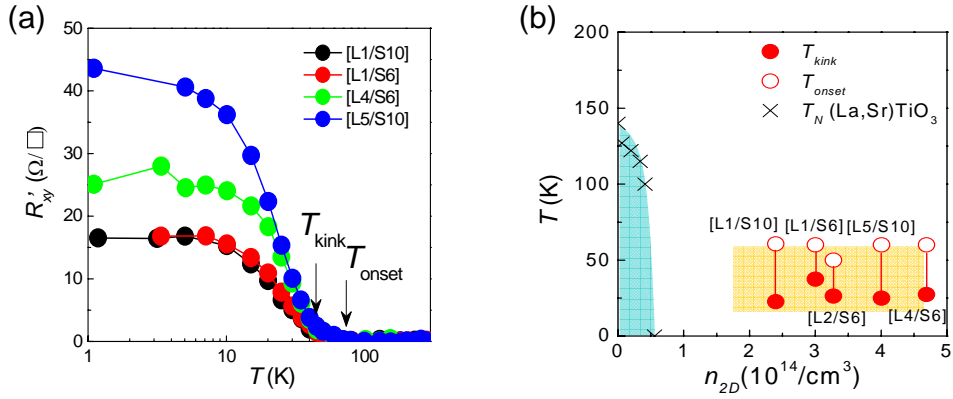

Figure S3. The carrier density dependence of the characteristic temperatures $T_{onset}$ and $T_{kink}$. At $T_{onset}$, $R'_{xy}(T)$ starts increasing, while at $T_{kink}$ $R'_{xy}(T)$ curve shows a maximum change of the slope. For comparison, the Neel temperature ($T_N$) of the Sr doped LaTiO$_3$ as a function of the carrier density is presented together.

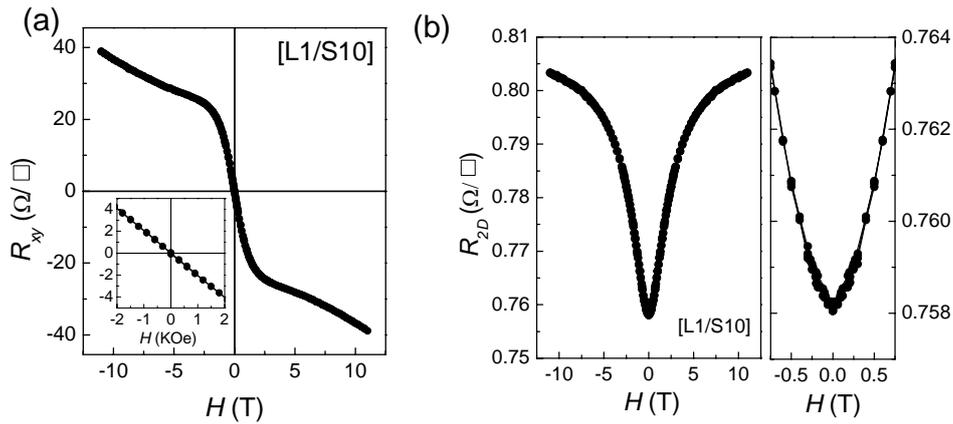

Figure S4. Magnetic field dependent Hall and sheet resistance, $R_{xy}(H)$ and $R_{2D}(H)$ at T = 1.1 K for (LaTiO$_3$)$_1$/(SrTiO$_3$)$_{10}$.

Considering that Brinkman *et al.* reported magnetic hysteresis for the LaAlO$_3$/SrTiO$_3$ interface at 0.3 K, we cannot completely rule out the possible observation of such a magneto-hysteresis in our LaTiO$_3$/SrTiO$_3$ superlattices at lower temperatures. However, since the strong non-linear behaviour in the Hall resistivity shows up below ~ 50 K, the absence of the magneto-hysteresis in this temperature range points towards a non-magnetic origin of the observed anomalous behaviour of the magnetotransport properties in the LaTiO$_3$/SrTiO$_3$ interface.

In the recent theoretical studies on the LaTiO$_3$/SrTiO$_3$ interface [4], the "Mottness" of LaTiO$_3$ is readily destroyed by the electronic reconstruction leading to formation of additional holes in LaTiO$_3$. Such a leakage of the Ti $d^1$ electronic wave function into the SrTiO$_3$ side is expected to substantially reduce the magnitude of the Coulomb potential ($U$) compared to its bulk value, diminishing the tendency towards magnetism, as observed in doped bulk LaTiO$_3$. These findings, therefore, strongly suggest that another mechanism is responsible for the nonlinear magnetotransport properties of LaTiO$_3$/SrTiO$_3$ interfaces.

## 3. Multichannel conduction in LaTiO$_3$/ SrTiO$_3$ superlattices

As an origin of the strongly nonlinear Hall effect, we consider a multi-channel conduction mechanism involving different subbands in the V-shaped potential well due to the delta doping. The two-carrier model is used to fit the nonlinear Hall data as shown in Fig. 2 and Fig. S3 (solid lines). In this model, the contribution from each type of carriers can be described by magnetoconductivity tensor $\sigma^i$ with elements $\sigma^i_{xx} = \sigma^i_{yy} = \rho_i/[\rho_i^2 + (R_iB)^2]$ and $\sigma^i_{xy} = -\sigma^i_{yx} = -R_iB/[\rho_i^2 + (R_iB)^2]$, where $\rho_i = 1/n_ie\mu_i$.[1] The sum of the contribution from all the carriers is simply the total conductivity $\sigma = \sum \sigma^i$, which is inverse of the total resistivity, *i.e.* $\rho = \sigma^{-1}$. Here $\mu_i$ is the mobility of carriers and $n_i$ is the carrier density (negative for electrons and positive for holes). When only two contributions to conduction are taken into account, the equations for the Hall resistance is generally written as

$R_H = [(\mu_1^2n_1+\mu_2^2n_2) + (\mu_1\mu_2)^2(n_1+n_2)]/e[(\mu_1|n_1| + \mu_2|n_2|)^2 + (\mu_1\mu_2B)^2(n_1 + n_2)^2]$ : (1)

Based on this equation, we fit $R_{xy}(H)$ data with a constraint of $R_{xx}(0) = 1/e(n_1\mu_1+n_2\mu_2)$. The fitted parameters are plotted in Fig. 2 and Fig. S3. The magnitude and the detailed behaviour of the Hall resistivity with magnetic fields are somewhat different from each other depending on the superlattice configuration, which suggests that the Hall resistivity is very sensitive to the subtle change of the charge transport such as the impurity scattering at the interface. However, the fitted parameters for several samples fall into a narrow range of values, and thus the multi-carrier transport is indeed a common feature in the LaTiO$_3$/SrTiO$_3$ superlattices.

There can be several other possible descriptions, rather than the broad charge distribution near the delta-doping as proposed in the text, for the multi-carrier transport in LaTiO$_3$/SrTiO$_3$ superlattice. First, the charge transfer between the LaTiO$_3$ and the SrTiO$_3$ layers can generate two types of carries in each side, which concurrently contribute to conduction. In this case, we can expect that there exists a comparable amount of holes and electrons in LaTiO$_3$ and SrTiO$_3$ layers, respectively. However, our experimental findings are that both majority and minority carriers are electrons and also their densities are significantly different from each other as shown in Fig. 3. Secondly, in addition to the interface-induced carriers, there might be charge carriers generated by oxygen defects in SrTiO$_3$ during the film growth, as intensively discussed for the LaAlO$_3$/SrTiO$_3$ interfaces reported [5,6]. However, as mentioned before, our growth condition rules out such an extrinsic effect, as also confirmed by the results in Fig. 2 (a) and (b). Therefore, we assume that the charge carrier induced by the oxygen defects is negligible, and, if any, it should be ascribed to the minority carriers rather than the majority carriers. For the carriers arising from oxygen defects, they reside in narrow impurity bands crossing the Fermi level and become mobile. As temperature decreases, therefore the impurity band becomes so narrow that it no longer intersects the Fermi level, leading to a strong reduction in the carrier density at low temperatures, often called "freezing-out". [8] However, our experimental data does not exhibit such a freezing-out behaviour, but rather show an opposite trend (see Fig. 3(a)); the minority carrier density rapidly increases with decrease of temperature. Therefore, this hypothesis also has to be ruled out.

**4. Electric field effect on the multichannel conduction in LaTiO$_3$/ SrTiO$_3$ superlattices**

The magnetic field dependent Hall resistivity of the [L4/S6] superlattices at 1.6 K with different gate voltages is shown in Fig. S5. [9] As found in the [L2/S6] superlattices (see Fig. 4 in the text), the non-linear dependence on the magnetic filed is significantly changed by the applied electric fields. At high negative gate voltages, the Hall resistivity shows a conventional linear dependence, while it is converted to the strongly non-linear behaviour at high positive gate voltages. As described in the text, we have used the two carrier model for fitting the data, indicated by the solid line in the figure. The obtained fitting parameters are plotted in Fig. S4(b) and (c). Consistent with the results of the [L2/S6] superlattices, the carrier density of the majority carriers and its mobility shows a moderate dependence on the gate voltage, while for the minority carriers, its density and mobility are strongly modified by the gate voltage. At negative gate voltages, the density of minority carriers is negligible or under the detection limit of our techniques. However, as the gate voltage increases, the minority carrier density is enhanced rapidly and then saturated at higher gate voltages.

Such a complex behaviour is in contrast to the conventional behaviour in the semiconductor field effect device where the applied electric field through the gate dielectric layer simply attracts or repulses the charge carrier to the interface modifying the carrier density of the semiconductor. First, the dielectric constant ($\varepsilon$) of SrTiO$_3$ shows strong electric field dependence. [6] As the applied voltage increases, $\varepsilon$ of SrTiO$_3$ is rapidly reduced, thus the resulting electric field increases more and more slowly with increasing applied gate voltage. This is in fact related to the saturation of the carrier density at higher gate voltage as shown in Fig. 4(d) and Fig. S5(b).

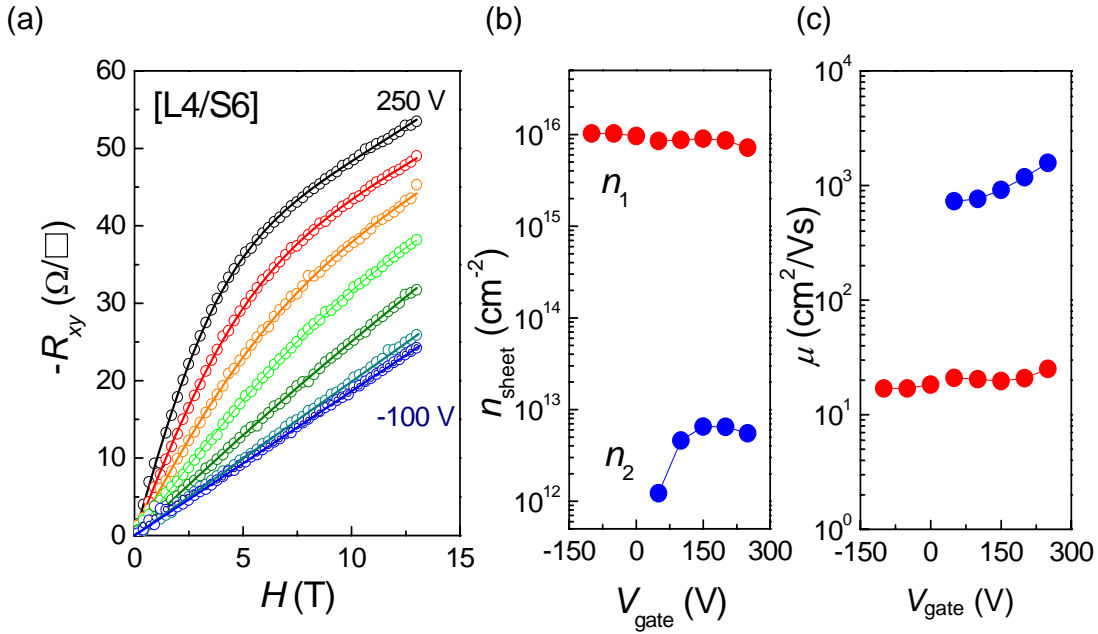

Figure S5. (a) Hall resistance of a (LaTiO$_3$)$_4$/(SrTiO$_3$)$_6$ superlattice as a function of magnetic fields for various gate voltages from 250 to -100 V at a 50 V step. The solid lines are fitted curves using the two-carrier models described in the text. The gate voltage dependence of (b) the carrier densities and (c) the mobilities for two types of charge carriers.

Furthermore, in our LaTiO$_3$/SrTiO$_3$ superlattices, the dielectric SrTiO$_3$ substrate itself also participates in the charge transport by providing a conducting channel near the interface with the first LaTiO$_3$ layer due to the band-bending effect. The applied electric field, therefore, modifies the potential profile in the first SrTiO$_3$ layer as illustrated in the inset of Fig. 4(c). This effect is essential to control the spatial extent of the electron gas, as discussed in the text, which effectively modifies the densities and the mobilities of each carrier in the multi-carrier transport.

In order to understand the detailed behaviours of our LaTiO$_3$/SrTiO$_3$ superlattices under the applied electric field, we need to understand how wide the charge carriers can be spread especially at low temperatures. Note that our superlattices consist of 10-20 LaTiO$_3$ layers embedded in SrTiO$_3$. In this configuration, the conducting channels formed at each LaTiO$_3$/SrTiO$_3$ interface contribute to the charge transport. The separation between two neighbouring interfaces is typically 2–4 nm (6–10 u.c. of $SrTiO_3$) in our samples, thus they are well-separated at room temperature, considering that the spread of the induced electrons is ~ 1 nm. [10] In this case, due to the relatively small dielectric constant of $SrTiO_3$ ($\varepsilon_r$ ~ 300), [11] the electron wave function is confined near the V-shaped potential wells for all delta doped layers. At low temperatures, however, enhanced screening of scattering canters by help of the high dielectric constant of $SrTiO_3$ $\varepsilon_r$ ~ 20,000 [11] leads to a significant spread of the electron wavefunction on the order of 10 nm, as also confirmed in $LaAlO_3/SrTiO_3$ heterostructures. [12] Therefore, while the most of electron density remains confined near the delta doped layers, its tails begin to overlap with each other as well as with the scattering centers in the neighboring interfaces. As a consequence, in our superlattice geometry, the minority carriers would experience large scattering due to the increased interactions with carriers from neighboring interfaces. The first interface from the substrate is however an exception as demonstrated in the Fig. S6. The tail of electron wavefunction at the first interface can penetrate deep into the $SrTiO_3$ substrate, thus keeping sufficient distance from the scattering centers. The most of minority carriers with high mobility in our superlattice samples, therefore, are expected to reside close to the first interface with the substrate, contributing dominantly to the transport properties.

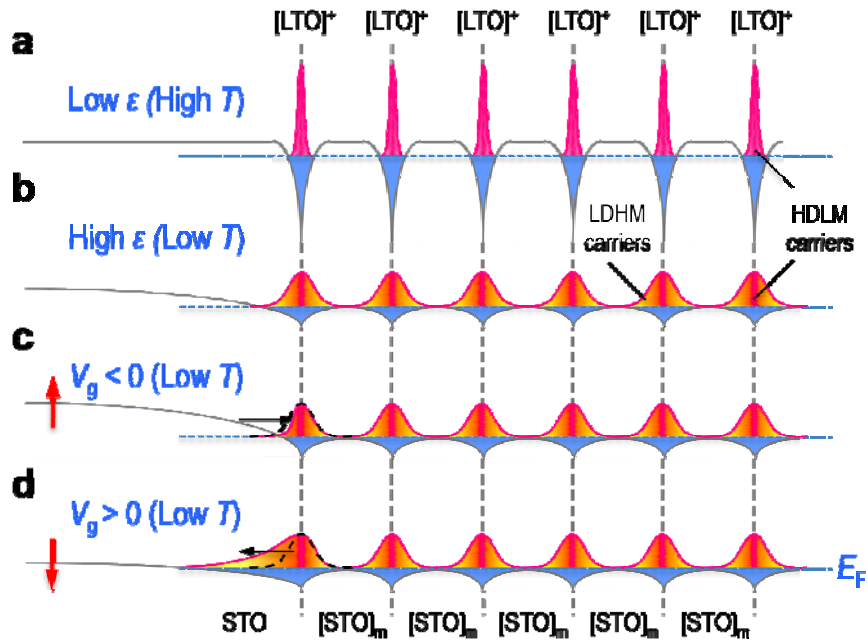

Figure S6. Schematic potential profiles and wave functions of charge carriers for a LaTiO$_3$/SrTiO$_3$ superlattice. (a), (b) The relatively low (high) dielectric constant $\varepsilon$ of SrTiO$_3$ at high (low) temperatures makes the narrow (wide) spread of the electron density profile. The regions with strong carrier scattering are illustrated in orange. (c), (d) The effects of external electric field on the potential profile and wave function of charge carriers in the superlattice at low temperatures. Here the gate electrode is assumed to be placed on the left-hand side of the delta-doped layer. Note that due to the thin SrTiO$_3$ layers within the superlattices reported here the low density carriers generated within the superlattices still suffer from strong scattering by high density carriers at the interfaces, except for the SrTiO$_3$ substrate side. While the high mobility carriers can be dominant within the superlattice structure when the SrTiO$_3$ layers are thick enough, the gating experiment influences only the carrier profile of the first interface due to the strong screening of applied electric field by the high density carriers (*i.e.,* 0.5$e$/interface unit).

This hypothesis can also explain the electric-field effect on the transport properties of our superlattice. When the gate voltage is applied, the electric field affects mostly the charge carriers at the first interface from the substrate, *i.e.* the closest interface to the gate electrode. The large charge carrier concentration of ~ $10^{14}$ $cm^{-2}$ each interface is expected to be very effective to screen the electric field. The charge carriers in the other interfaces, therefore, feel much less electric field, and their potential profiles are almost intact. However the potential profile near the first interfaces is significantly modified, and as a result, the applied electric field mostly affects the minority carrier near the first interface. At high negative voltages, therefore, the highly mobile minority carriers are almost completely suppressed as shown in Fig. 4 and Fig. S5, supporting this hypothesis.

As for the majority carriers, the total density of the majority carriers, $n$ estimated from Hall resistivity data, is determined by the majority carrier density at the first interface from the bottom, $n_1$ and the averaged majority carrier at the other interfaces $n_{other}$ based on the equation, [13]

$$n = (n_1\mu_1 + n_{other}\mu_{other})^2/(n_1\mu_1^2 + n_{other}\mu_{other}^2).$$
With $\mu_1/\mu_{other} = \beta$,
$$n = (n_1\beta + n_{other})^2/(n_1\beta^2 + n_{other}).$$

Since the attracted carriers due to the positive $V_g$ are mostly located at the first interface, the carrier density at the first interface $n_1$ increases, while $n_{other}$ can be reduced with a positive $V_g$. Usually the mobility is inversely proportional to the carrier density, so $\beta$ in the equation above will be reduced with increasing the gate voltage. If the gate voltage dependence of $\beta$ is larger than that of $n_1$, the effective total carrier density $n$ can be reduced by increasing the gate voltage. This is opposite to what is expected in the simple picture of the conventional field effect device. These considerations can be a possible origin for the slightly decreasing behaviour of the majority carrier with $V_g$ in Fig. 4 (d) and Fig. S5(b).